\documentclass{ws-ijmpd}
\usepackage[super,compress]{cite}
\usepackage{graphicx}
\usepackage{subfigure}
\usepackage{amsmath}
\usepackage{amssymb}
\usepackage{wrapfig}
\usepackage{color}
\usepackage[english]{babel}
\usepackage[warning,math]{easyeqn}
\usepackage[thinlines]{easybmat}
\usepackage{color}

\begin{document}

\markboth{Yves-Henri Sanejouand}
{A FRAMEWORK FOR THE NEXT GENERATION \\ 
OF STATIONARY COSMOLOGICAL MODELS}

\catchline{}{}{}{}{}

\title{
A FRAMEWORK FOR THE NEXT GENERATION \\ 
OF STATIONARY COSMOLOGICAL MODELS}

\author{YVES-HENRI SANEJOUAND}

\address{Facult\'e des Sciences et des Techniques, Nantes, France
\\
yves-henri.sanejouand@univ-nantes.fr}

\maketitle

\begin{history}
\received{January 2022, 4$^{th}$.}
\revised{April 2022, 12$^{th}$.}
\revised{June 2022, 4$^{th}$.}
\end{history}

\begin{abstract}
According to a new tired-light cosmological model, where $H(z)=H_0 (1 + z )$, the number density of galaxies has been nearly constant over the last 10 Gyr, at least, 
meaning that, as far as galaxy counts are concerned, the Universe has been stationary.  
In this context, an analysis of the luminosity distances of quasars and supernovae Ia shows that the Universe is far from being as transparent as assumed nowadays, the photon lifetime along the line-of-sight being one third of the Hubble time.   
Such a "low" value could mean that there are huge amounts of grey dust in the inter galactic medium, that have so far escaped detection. It could also be a signature of "new physics", namely, a consequence of the decay of photons into lighter particles. 

The tired-light model advocated in the present study would be falsified if, for instance, the time-dilation of remote events were shown to have a general character, that is, if it were definitely observed for phenomenons other than the light curves of supernovae Ia. 
On the other hand, further developments are needed in order to turn this model into an actual cosmology.
In particular, the physical origin of the current stability of the Universe on galactic scales remains to be identified.

\end{abstract}


\ccode{PACS numbers: 98.80.-k}

\section{Introduction}

The family of cosmologies initiated by Georges Lema\^itre \cite{Lemaitre:27} proved able to make challenging predictions. Among them: 
luminosity distances are larger than angular ones by a factor of $( 1 + z )^2$;  
\textit{all} remote events look slower than local ones by a factor of $( 1 + z )$;
there is an isotropic radiation with the spectrum of a blackbody at a temperature of $T_0 ( 1 + z )$, $T_0$ being its local temperature \cite{Tolman:30,Wilson:39,Herman:49}.

Such predictions have been backed by numerous observations. For instance, the expected time-dilation of remote events has been found in the 
light curves of supernovae Ia \cite{Perlmutter:96,Riess:96,Blondin:08}, a thermal radiation at a temperature of $T_0$ = 2.73 K has been observed \cite{Penzias:65,Dicke:65} and its redshift dependence has been confirmed \cite{Zenteno:14,Lamagna:15,Lima:16}.

Still, several clouds are obscuring the brilliance of $\Lambda$CDM \cite{Ostriker:95,Krauss:95,Peebles:03}, 
the so-called "concordance cosmology" \cite{Tegmark:01}.
Among them, the Hubble tension seems to attract most of the attention \cite{Silk:21}.
However, other clouds may prove darker \cite{Lopez:17,Skara:21}. 
In particular, $\Lambda$CDM is based on a "cosmic trinity" \cite{Silk:20} of three essential ingredients with weird properties and for which there is no direct evidence, namely, an early stage of accelerated expansion \cite{Starobinsky:80,Guth:81,Starobinsky:82,Linde:82}, 
a dark matter and an energy components of unknown nature \cite{Peebles:03,Bartelmann:10}, both accounting for $\approx$ 95\% of the matter-energy content of the Universe \cite{Planck:18}.

In other words, according to $\Lambda$CDM, the dominant forms of matter-energy are of a different nature on Earth and far away.
Though such an hypothesis was taken for granted in the ancient times, it is the opposite hypothesis that has proven fruitful since the Renaissance, namely, that what is observed on Earth is representative of what is found in the rest of the Universe. 
Given the numerous successes of the later hypothesis, it seems reasonable to push it forward once more.  

However, it is not obvious to build from scratch a new cosmology able to compete with  
the result of the work of several generations of brilliant scientists. 
Also, given the huge time and distance scales involved in cosmological problems, key physical phenomenons may still be not accounted for, like a tiny variation of quantities nowadays assumed to be constant\cite{Dirac:37,Webb:99,Sanejouand:09}. 

So, as a preliminary step, it may prove useful to pinpoint a set of ingredients which could serve as a basis for the development of an alternative family of cosmological models. 
To do so, a consistent set of simple relationships able to handle a large range of observables has been looked for. As a starting point, a tired-light model was chosen, mostly because such models have not been scrutinized as much as models based on metric theories of gravity.

\section{Main hypothesis}

\subsection{A generic tired-light model}

As proposed a while ago \cite{Tetrode:22,Feynman:45}, let us assume that photons can not fly away for ever. However, instead of interpreting electromagnetic radiation as an interaction between a source and an absorber, let us posit that photons have \textit{all} the same maximum range, $d_H$, due to a loss of their energy such that:
\begin{equation}
\label{eq:linear}
h \nu_{obs} = h \nu_0 - f_\gamma d_T
\end{equation}
where $\nu_0$ is the frequency of the photon when it is emitted, $\nu_{obs}$, its frequency when it is observed, $d_T$, the distance between its source and the observer, $h$ being the Planck constant.

$\nu_{obs} = 0$ when $d_T = d_H$. So,
$f_\gamma = \frac{h \nu_0}{d_H}$ and
eqn \ref{eq:linear} can also be written as follows:
$$
\nu_{obs} = \nu_0 ( 1 - \frac{d_T}{d_H} )
$$
that is:
\begin{equation}
\label{eq:maxhubble}
\frac{z}{1+z} = \frac{d_T}{d_H} 
\end{equation}
Thus, when $z \ll 1$:
$$
z \approx \frac{d_T}{d_H} 
$$
So, assuming that:
\begin{equation}
\label{eq:range}
d_H=\frac{c_0}{H_0}
\end{equation}
yields:
$$
z \approx \frac{H_0}{c_0} d_T
$$
which is the relationship anticipated by Lema\^itre \cite{Lemaitre:27} and further confirmed by Hubble \cite{Hubble:29}, $H_0$ being the Hubble constant, $c_0$, the speed of light. 

Though the idea that the Hubble-Lema\^itre law is the result of some tired-light mechanism has been proposed a number of times,\cite{Zwicky:29,Nernst:37,Finlay:54,deBroglie:66,Vigier:88,Heymann:14} note that the hypothesis that photons may \textit{all} have same range has been, to my knowledge, little considered so far. 
Indeed, most tired-light mechanisms proposed previously
yield an exponential decay of the photon energy\cite{North,Marmet:18},
in line with the classic hypothesis that the photon 
has an infinite range.    
  
\subsection{Consistency with H(z) data}

\begin{figure}[t]
\includegraphics[width=7.5 cm]{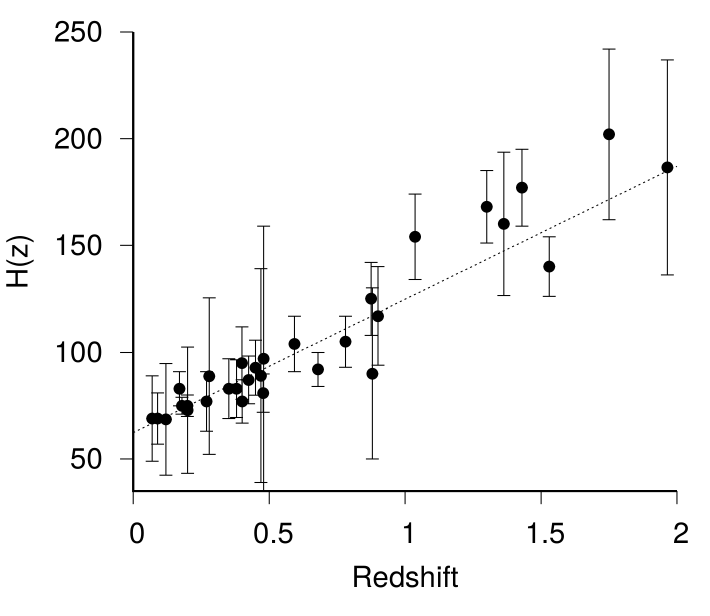}
\vskip -0.5 cm
\caption[]{
H(z) as a function of redshift, as obtained with the cosmic chronometer method.
Dotted line: weighted linear fit. 
}
\label{Fig:hofz}
\end{figure}

Let us now assume that, as postulated by Einstein \cite{Einstein:05rr}, and as checked in various contexts\cite{Will:90,Schaefer:99,Schiller:09,Kostinski:12}, the speed of light is constant and that delays, 
due for instance to the Shapiro effect \cite{Shapiro:79}, can be neglected in such a way that:
\begin{equation}
\label{eq:cste}
d_T \approx c_0 \Delta t 
\end{equation}
where $\Delta t = t_0 - t$ is the photon time-of-flight, $t_0$ and $t$ being the observer and cosmic times, respectively.
Thus, with eqn \ref{eq:range}, eqn \ref{eq:maxhubble} becomes:
\begin{equation}
\label{eq:newhubble}
\frac{z}{1+z} = \frac{H_0}{c_0} d_T 
\end{equation}
that is, with eqn \ref{eq:cste} \cite{Sanejouand:14}:
\begin{equation}
\label{eq:newhubblet}
\frac{z}{1+z} = H_0 \Delta t 
\end{equation}
As a consequence:
\begin{equation}
\frac{\partial z}{\partial t} = - H_0 (1 + z)^2
\label{eq:dzdt}
\end{equation}

Values of $\frac{\partial z}{\partial t}$ can be determined with the cosmic chronometer method, where $\partial t$ is estimated by measuring the age of passively evolving galaxies~\cite{Stern:10,Jimenez:12}. They are usually provided through $H(z)$, which is defined as follows~\cite{Jimenez:02}:
\begin{equation*}
H(z)= -\frac{1}{1 + z}\frac{\partial z}{\partial t}
\end{equation*}
So, with eqn~\ref{eq:dzdt}:
\begin{equation}
H(z) = H_0 ( 1 + z ) 
\label{eq:hz}
\end{equation}

It is indeed well known that, as illustrated in Figure \ref{Fig:hofz}, $H(z)$ data \cite{Wang:18} are consistent with a linear relationship \cite{Sanejouand:14,Kumar:12}.
As a matter of fact, eqn \ref{eq:hz} is also a prediction of linear 
coasting cosmologies \cite{Kolb:89,Walker:99,Chardin:12}, like the $R_h = c t$ cosmology developed by Fulvio Melia and his collaborators \cite{Melia:12} who have claimed that, compared to $\Lambda$CDM, it is favored by various model selection criteria \cite{Melia:13,Melia:18}. 

Note however that eqn \ref{eq:newhubblet} or \ref{eq:hz} have also been obtained in other physical contexts \cite{Heymann:14,Sanejouand:18}. 

\section{Counts of galaxies}

$n(d_T)$, the cumulative count of galaxies as a function of the light-travel distance, is such that:
\begin{equation}
n(d_T) = \int_0^{d_T} 4 \pi N(r) r^2 dr
\label{eq:count_r}
\end{equation}
where $N(r)$ is the number density of galaxies at distance $r$.

Let us assume that $N(\Delta t)$, the number density of galaxies as a function of the photon time-of-flight, evolves slowly enough, so that:
\begin{equation}
N(\Delta t) \approx N_G + \dot{N} \Delta t
\label{eq:dens_slow}
\end{equation}
where $N_G$ is the local number density, $\dot{N}$ being the time derivative of $N(\Delta t)$.
With eqn \ref{eq:cste} and \ref{eq:dens_slow}, eqn \ref{eq:count_r} yields:
\begin{equation}
n(d_T) = \frac{4}{3} \pi d_T^3 N_G \left( 1 + \frac{3}{4} \frac{\dot{N}}{N_0} \frac{d_T}{c_0} \right)
\label{eq:count_d}
\end{equation}
which becomes, with eqn \ref{eq:range} and \ref{eq:newhubble}:
\begin{equation}
n(z) = n_{st} \frac{z^3}{(1+z)^3} \left( 1 + n_z \frac{z}{1+z} \right) 
\label{eq:count_z}
\end{equation}
where:
\begin{equation}
n_{st} = \frac{4}{3} \pi d_H^3 N_G
\label{eq:ngaltot}
\end{equation}
and:
$$
n_z = \frac{3}{4} \frac{\dot{N}}{N_G} H_0^{-1} 
$$

\begin{figure}[t]
\vskip  0.2 cm
\hskip -0.1 cm
\includegraphics[width=8.0 cm]{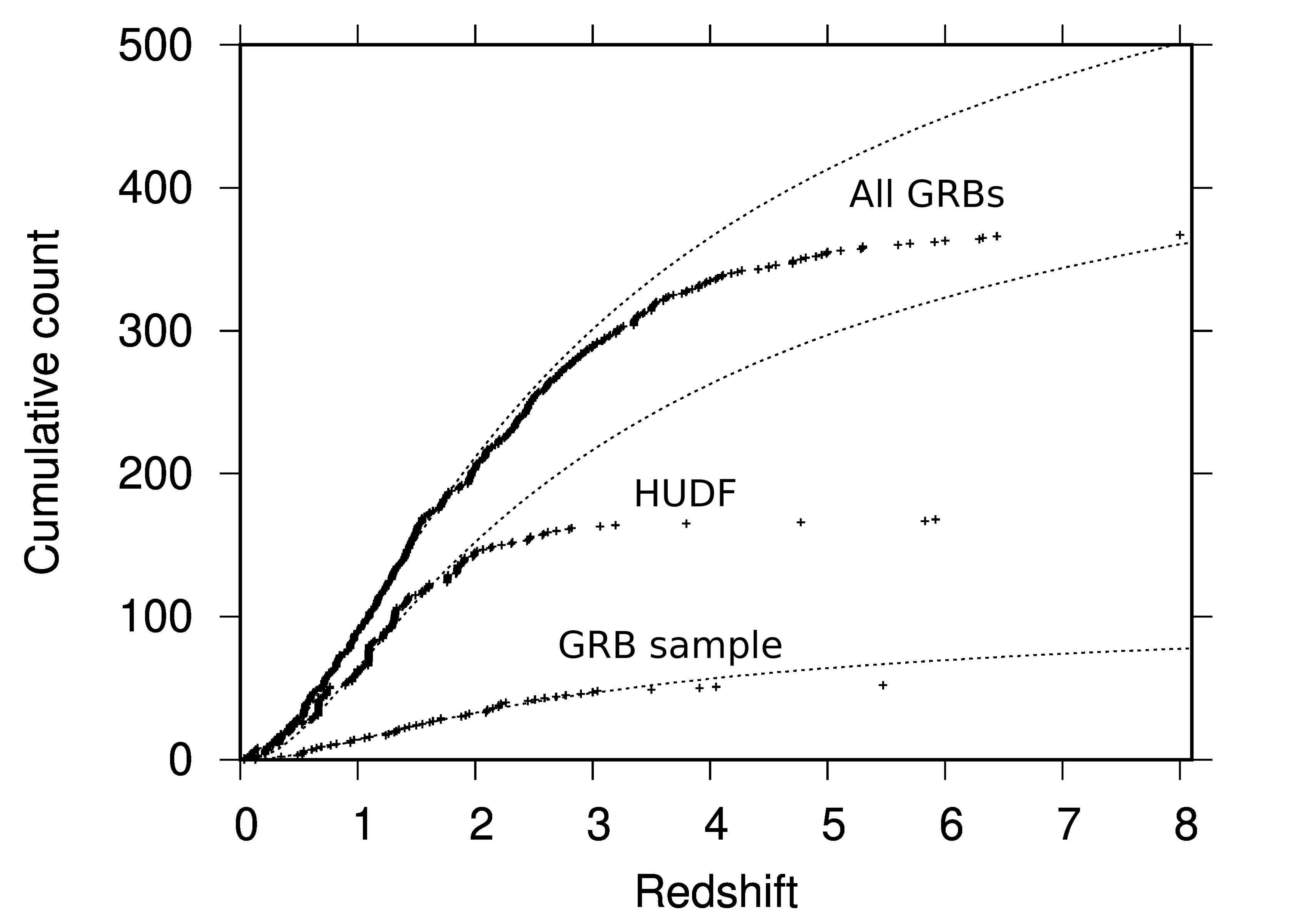}
\vskip -0.3 cm
\caption{Cumulative counts of galaxies as a function of redshift. 
Top and bottom: sources of long gamma-ray bursts (GRB) with a low (top) or high (bottom) redshift completeness level. Middle: galaxies in the Hubble Ultra Deep Field (HUDF), with robust spectroscopic redshifts. Dotted lines: single-parameter fits, the corresponding number density of galaxies being assumed constant for $z \leq$ 2.    
}
\label{Fig:count}
\end{figure}

In order to estimate $n_z$, that is, to assess the evolution of the number density of galaxies as a function of redshift, two datasets of galaxies with a fair level of redshift completeness were chosen, namely, the sources of gamma-ray bursts (GRBs) found by \textit{Swift}
and the ensemble of galaxies in the Hubble ultra deep field (HUDF)
for which a robust spectroscopic redshift has been obtained.    

Least-square fitting, for $z \leq 2$, of the cumulative count of 52 \textit{Swift} GRBs from a carefully selected sub-sample with a redshift completeness level of 90\% \cite{Tagliaferri:12} yields $n_z$ = -0.13 $\pm$ 0.10 ($n_{st}$ = 120 $\pm$ 8), confirming that the evolution of the number density of GRBs has been slow, with respect to the Hubble time ($H_0^{-1}$), as assumed above (eqn \ref{eq:dens_slow}). 
As a matter of fact, if the number density of GRBs does not vary as a function of redshift ($\dot{N}$ = 0, that is, $n_z = 0$), the root-mean-square of the residuals is 0.80 ($n_{st}$ = 111 $\pm$ 1), instead of 0.79, meaning that both fits are equally consistent with observational data.
When the 368 \textit{Swift} long ($t_{90} \geq$ 0.8 s)\cite{Sari:13} GRBs with a redshift known with fair accuracy are considered,\footnote{As provided on the Neil Gehrels Swift Observatory web page, (https://swift.gsfc.nasa.gov/archive/grb\_table), on August 2021, 12$^{th}$.} the root-mean-square of the residuals is higher, namely 4.0 ($n_{st}$ = 713 $\pm$ 2), maybe as a consequence of the much lower redshift completeness level (redshifts are known for only 30\% of the GRBs detected by \textit{Swift} \cite{Swift}).

On the other hand, fitting the cumulative count of the 169 galaxies in the HUDF with robust spectroscopic redshifts\footnote{As found in Table 4 of Rafelski and col.\cite{Voyer:15}.}, for $z \leq 2$, yields a similar root-mean-square of the residuals, namely, of 5.5 ($n_{st}$ = 513 $\pm$ 3). 

So, as shown in Figure \ref{Fig:count}, when the number density of galaxies is assumed to be constant ($n_z$ = 0), eqn \ref{eq:count_z} allows for a fair single-parameter fit of the observational data, up to $z \approx 2$ at least. 
Note that the mismatch observed for $z > 2$ could be due to the difficulty of redshift determination at such distances.

Since long GRBs occur in star-forming galaxies \cite{Wiersema:12,Tagliaferri:16}, the fact that the number density of GRB sources does not vary significantly for $z < 2$ suggests that the number density of star-forming galaxies does not as well, as already indicated by previous studies \cite{Wisotzki:06,Williams:11}. 
On the other hand, the fact that the number density of galaxies is also found nearly constant in the HUDF further suggests that the number density of quiescent galaxies follows the same trend. 

So, according to the above analysis, as far as galaxy number densities are concerned, the Universe seems to have been stationary over the last 10 Gyr ($z < 2$), at least. 

Note that, with $\Lambda$CDM as the background cosmology, the star-formation rate density is instead assumed to have a complex history, namely, to have peaked approximately 3.5 Gyr after the Big Bang, and declined exponentially at later times \cite{Dickinson:14}. 

\section{Angular distance}

Stationarity on galactic scales could mean that,
as suggested in a number of previous studies\cite{Einstein:17,Laviolette:86,Crawford:93,Scarpa:14,Lerner:18},
the space-time metric of the Universe is static.
 
If so, $d_A = d_T$, $d_A$ being the angular distance, and $\theta_s$, the angular size of a remote object, is so that:
$$
\theta_s \approx \frac{s}{d_T}
$$
where $s$ is the actual size of the object. That is, with eqn \ref{eq:maxhubble} \cite{Sanejouand:14}:
$$
\theta_s \approx \frac{s}{d_H} ( 1 + \frac{1}{z} )
$$
Thus, for $z \gg 1$, $\theta_s \rightarrow \theta_{min}$, with:
\begin{equation}
\theta_{min} = \frac{s}{d_H} 
\label{eq:theta}
\end{equation}
Interestingly, measurements of the average angular size of Lyman-break galaxies are indeed consistent with constant angular sizes above $z \approx 2.5$, with a mean half-light radius of 0.24$''$ at $z \approx 4$ \cite{Stern:04}.
  
According to eqn \ref{eq:range} and \ref{eq:theta}, with $H_0 \approx$ 70 km$\cdot$s$^{-1}\cdot$Mpc$^{-1}$
\cite{Planck:18,Riess:16,Paturel:17,Meylan:20}, these results mean that the average light radius of Lyman-break galaxies is $s \approx$ 8 kpc, that is, the order of magnitude of the optical radius of a star-forming galaxy in the local Universe.
They also mean that the average linear size of Lyman-break galaxies has not changed significantly over the last 12 Gyr ($z < 4$).

On the contrary, with $\Lambda$CDM as the background cosmology,
the average linear size of galaxies, $\langle r_G \rangle$,
is assumed to experience a strong evolution, with
$\langle r_G \rangle \propto (1 + z )^n$,
$n$ ranging between -1.4 and -0.8 \cite{Vandokkum:17,Nakajima:19},
$\langle r_G \rangle$ being for instance six times smaller at z = 3.2 than at z = 0 \cite{Lopez:10}.

\section{Luminosity distance}

In a static Universe, $d_L$, the luminosity distance, is expected to have a form like:
\begin{equation}
d_L = d_T (1+z)^{\frac{1}{2}} e^{\frac{1}{2} \tau(z) }
\label{eq:Dlgenerald}
\end{equation}
where the $(1+z)^{\frac{1}{2}}$ term corresponds to the energy loss of the photons during their travel, while $\tau(z)$ denotes the opacity between the source and the 
observer \cite{Holanda:14}, herein assumed   
to be for the most part due to a single physical phenomenon. As such, it may prove well 
described by the Beer-Lambert law, 
so that the optical depth can be parametrized as:
\begin{equation}
\tau(z) = \frac{\varepsilon}{d_H} d_T
\label{eq:opacity}
\end{equation}
that is, with eqn \ref{eq:maxhubble}: 
$$
\tau(z) = \varepsilon \frac{z}{1+z}
$$
a parametrization that has also been considered in the context of metric theories of gravity\cite{Dantas:17}. Then,  
eqn \ref{eq:Dlgenerald} becomes:
\begin{equation}
d_L = d_H \frac{z}{\sqrt{1+z}} e^{\frac{\varepsilon}{2} \frac{z}{1+z}}
\label{eq:Dlgeneral}
\end{equation}

\subsection{Distance modulus}

With eqn \ref{eq:Dlgeneral}, $\mu$, the distance modulus:\footnote{$d_L$ being in Mpc.}
$$
\mu = 5 \log_{10}( d_L ) + 25
$$    
becomes:
\begin{equation}
\mu = 5 \log_{10} \frac{z}{\sqrt{1+z}} + \alpha \varepsilon \frac{z}{1+z} + \mu_0
\label{eq:mu}
\end{equation}
where $\mu_0=5 \log_{10} d_H + 25$, while $\alpha= 2.5 \log_{10} e$.

\subsection{Quasars}

\begin{figure}[t]
\hskip -0.3 cm
\includegraphics[width=8.0 cm]{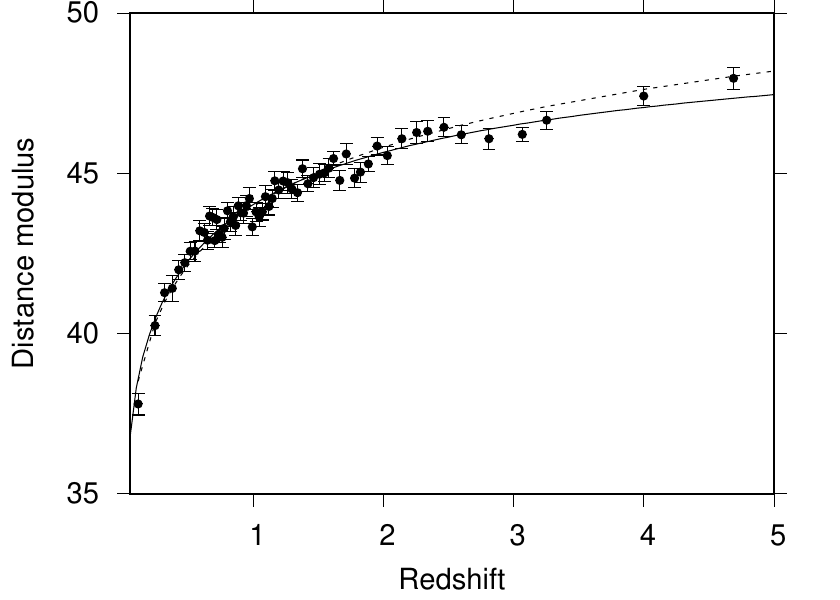}
\vskip -0.3 cm
\caption[]{The distance modulus of an homogeneous sample of quasars, as a function of their redshift. Each point (filled circles) is an average over 25 quasars, error bars indicating the corresponding standard errors. Dashed line: $\Lambda$CDM. Plain line: our two-parameter weighted least-square fit.
}
\label{Fig:quasars}
\end{figure}

In order to estimate $\varepsilon$, a homogeneous sample of 1598 quasars \cite{Lusso:19} with a distance modulus determined using their rest-frame X-ray and UV fluxes \cite{Lusso:15,Risaliti:16} was considered. 
To get accurate average values from these distance moduli,
the dataset was sorted by increasing redshift values and split into 64 groups of 25 quasars with similar redshifts,\footnote{With 23 quasars in the highest-redshift group.} the average redshift and distance modulus of each group being used for the present analysis, 
together with the corresponding standard error.

A weighted least-square fit of these 64 average distance moduli yields 
$\varepsilon$ = 3.1 $\pm 0.3$
($\mu_0$ = 43.1 $\pm$ 0.2). 
Interestingly, as shown in Figure \ref{Fig:quasars}, eqn \ref{eq:mu} matches the observational data
(reduced $\chi^2_{dof}$ = 1.20) 
over the whole redshift range, namely, up to $z \approx$ 5.   

With $\Lambda$CDM as the background cosmology
($\Omega_m$ = 0.315, $\Omega_k$ = 0)\cite{Planck:18},
the overall quality of the fit 
is poorer
(reduced $\chi^2_{dof}$ = 1.47).
It can however be improved, for instance by allowing 
significantly larger values for $\Omega_m$ \cite{Ratra:20,Zhu:21}.

\subsection{The photon lifetime}

Let us now consider the hypothesis that the lifetime of photons along the line-of-sight is for the most part determined by their interactions with galaxies (or their halos). Thus, since, as found above, $\dot{N} \approx 0$, eqn \ref{eq:opacity} can also be written as follows: 
\begin{equation}
\tau(z) \approx \sigma_G N_0 d_T
\label{eq:sigma}
\end{equation}
where $\sigma_G$ is the average cross section of a galaxy, $N_0$ being the average galaxy number density.
On the other hand, according to eqn \ref{eq:ngaltot}:
$$
N_0 = \frac{n_{tot}}{\frac{4}{3} \pi d_H^3}
$$ 
where $n_{tot}$ is the total number of galaxies in the visible Universe.
Thus, with eqn \ref{eq:range}, \ref{eq:opacity} and \ref{eq:sigma}:
$$
\sigma_G = \frac{\varepsilon}{3} \frac{4 \pi d_H^2}{n_{tot}}
$$
and since, as found above, $\varepsilon \approx 3$:
\begin{equation}
\sigma_G \approx \frac{4 \pi d_H^2}{n_{tot}}
\label{eq:sigmasolid}
\end{equation}

Because there are $\approx$ 10,000 galaxies in the Hubble Extreme Deep Field (HXDF) \cite{XDF}, assuming that most of the galaxies in this small area have been captured, and also that the HXDF is a representative enough sample of the sky, a rough estimate can be proposed for $n_{tot}$, namely, 
$4~10^{11}$. 

Then, according to eqn \ref{eq:range} and \ref{eq:sigmasolid},
if $H_0 \approx$ 70 km$\cdot$s$^{-1}\cdot$Mpc$^{-1}$, $\sigma_G = 4 \pi r^2_G \approx 5~10^{41}$ m$^2$, with a corresponding radius $r_G \approx$
7 kpc, that is, the order of magnitude of the optical radius of a galaxy in the local Universe.

Note however that
the above value of $\sigma_G$ is expected to be a sum over all the objects involved in the process responsible for the loss of photons.

\section{Other checks}

\subsection{Supernovae Ia}

\begin{figure}[t]
\hskip -0.5 cm
\vskip  0.05 cm
\includegraphics[width=8.0 cm]{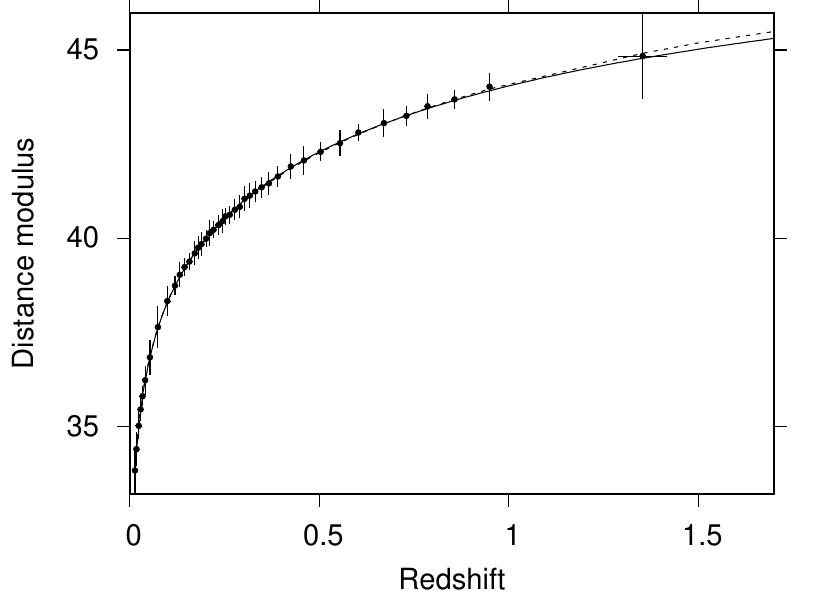}
\vskip -0.3 cm
\caption[]{The distance modulus of the supernovae Ia of the Pantheon sample, as a function of their redshift. Each point (filled circles) is an average over 25 SN Ia, error bars indicating the corresponding standard errors 
(multiplied by ten for the y-axis, for the sake of visibility).
Dashed line: $\Lambda$CDM. Plain line: as expected with $\varepsilon$ = 3.1.
}
\label{Fig:snIa}
\end{figure}

Though supernovae of type Ia (SN Ia) can not be studied over a range of redshifts as wide as quasars, their luminosity distance can be determined with a much higher accuracy \cite{Riess:98,Perlmutter:99}.
Like for quasars, the 1048 SN Ia of the Pantheon sample were gathered in 42 groups of 25 SN Ia with similar redshifts,\footnote{With 23 SN Ia in the highest-redshift group.} 
the average redshift and distance modulus of each group being used for the present analysis,
together with the corresponding standard errors.

As shown in Figure \ref{Fig:snIa}, with, 
as found above,
$\varepsilon$ = 3.1, 
the absolute magnitude of a SN Ia being $M=-19.36$ \cite{Union2:12sh} ($\mu_0$ = 43.1)\footnote{$\mu = m_B - M$, where $m_B$ is the apparent magnitude of the supernova.}, eqn \ref{eq:mu} matches the distance moduli of the SN Ia 
(reduced $\chi^2_{dof}$ = 1.14). 
Indeed, a weighted least-square fit of these data yields 
$\varepsilon$ = 3.15 $\pm$ 0.04 ($\mu_0 + M$ = 23.74 $\pm$ 0.01).

Note that with $\Lambda$CDM as the background cosmology ($\Omega_m$ = 0.315, $\Omega_k$ = 0),\cite{Planck:18} 
the quality of the fit is similar 
(reduced $\chi^2_{dof}$ = 1.09).

\subsection{Distance duality}

Let us write the cosmic distance duality relation as follows \cite{Holanda:10}:
\begin{equation}
d_L = \eta(z) d_A ( 1 + z )^2
\label{eq:duality}
\end{equation}

\begin{figure}[t]
\hskip -0.1 cm
\includegraphics[width=7.5 cm]{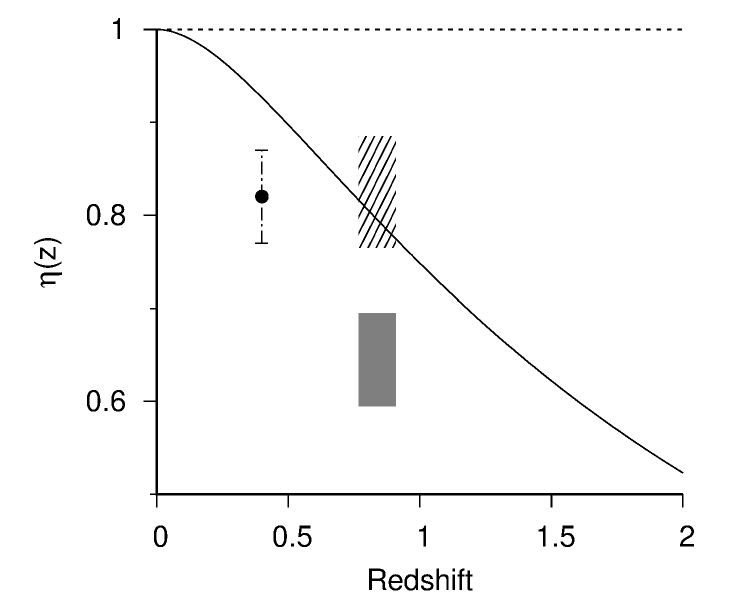}
\vskip -0.3 cm
\caption[]{Compatibility with the cosmic distance duality relation, as a function of redshift.
Boxes:
measurements obtained by studying early-type galaxies in the R (grey) and I (hatched) bands. Filled circle: in the K band.
Horizontal dashed line: \textit{minimum} value expected within the frame of metric theories of gravity, if there is no significant evolution of early-type galaxies.
Plain line: as expected 
with $\varepsilon$ = 3.1. 
}
\label{Fig:duality}
\end{figure}

In the case of this relationship, like most metric theories of gravity, $\Lambda$CDM predicts that $\eta(z) \geq 1$ \cite{Kunz:04}, with $\eta(z) = 1$ if there is no loss of photon along the path between the source and the observer \cite{Tolman:30}.

Interestingly, measurements of $\eta(z)$ tend to provide values that are below one \cite{Carvalho:96,Sandage:01,Uzan:04,Baryshev:08,Alcaniz:12,Holanda:16}. 
For instance, by studying 34 early-type galaxies from three clusters, with redshifts ranging between 0.72 and 0.92, Lubin and Sandage found $\eta(z)$ = 0.75--0.89,\footnote{Corresponding to an exponent on (1 + z) between 3.06 and 3.55 \cite{Sandage:01}.} in the I band, and $\eta(z)$ = 0.57--0.71,\footnote{Corresponding to an exponent on (1 + z) between 2.28 and 2.81 \cite{Sandage:01}.} in the R band \cite{Sandage:01}, in line with a previous value of $\eta(z)$ = 0.82 $\pm$ 0.05,\footnote{Corresponding to a surface brightness difference of 1.02 $\pm$ 0.14 mag.arcsec$^2$, when the k-correction is taken into account \cite{Carvalho:96}.} obtained by comparing the surface brightness of elliptical galaxies in the Abell 851 and Coma clusters,\footnote{At $z$ = 0.4 and 0.02, respectively.} in the K band \cite{Carvalho:96}.

With $\Lambda$CDM as the background cosmology, 
such values of $\eta(z)$ are explained by assuming a strong evolution of early-type galaxies \cite{Carvalho:96,Sandage:01,White:96,Holland:08}.

On the other hand, with eqn \ref{eq:newhubble} and \ref{eq:Dlgeneral}, eqn \ref{eq:duality} yields:
\begin{equation}
\eta(z) = (1+z)^{-\frac{3}{2}} e^{\frac{\varepsilon}{2} \frac{z}{1+z}}
\label{eq:eta_of_z}
\end{equation}

As shown in Figure \ref{Fig:duality}, according to eqn \ref{eq:eta_of_z}, with $\varepsilon$ = 3.1, 
for $z \gg 0$, 
$\eta(z)$ is indeed expected to have values that are well below one.

\subsection{Old high redshift objects}

Remote objects are observed like they were at time $t=t_0 - \Delta t$, where $\Delta t$ is the photon time-of-flight. On the other hand, the oldest objects in our neighborhood, like HD 140283, an extremely metal-deficient subgiant star nowadays known as the Methuselah of 
stars, or the globular cluster NGC 6101, are 13--14 Gyr old \cite{Bond:13,Turner:15,Chaboyer:17} while, 
for instance, the oldest object known 
at $z \approx 4$, namely, APM 08279+5255, an exceptionally luminous, gravitationaly lensed, quasar, seems to be 2--3 Gyr old \cite{Friaca:05,Komossa:02,Komossa:02x}. 

Such observations suggest that, as predicted by Lema\^itre cosmologies, the oldest objects known started to emit light approximately at the same time, at least $T_f \approx$ 13 Gyr ago. 
In the context of the present study, 
as far as the oldest objects are concerned, $T_{old}(z)$, their observed age at a given redshift, is expected to be: 
$$
T_{old}(z) = T_f - \Delta t
$$
that is, with eqn \ref{eq:newhubblet} \cite{Sanejouand:14}:
\begin{equation}
T_{old}(z) = T_f - \frac{1}{H_0} \frac{z}{1+z}
\label{eq:age}
\end{equation}
Interestingly, according to eqn \ref{eq:age}, if the age of old local objects like HD 140283 is assumed to provide a fair estimate for $T_f$, with $H_0 \approx$ 70 km$\cdot$s$^{-1}\cdot$Mpc$^{-1}$, the oldest objects at $z \approx 4$ are expected to have an age of $T_{old}(4) \approx 2.7$ Gyr, in good agreement with the measured age of APM 08279+5255.

Note that it has been underlined that $\Lambda$CDM can hardly cope with the age of APM 08279+5255 \cite{Friaca:05,Sethi:05,Dev:06,Zhang:10}.

\begin{figure}
\hskip -0.1 cm
\includegraphics[width=8.0 cm]{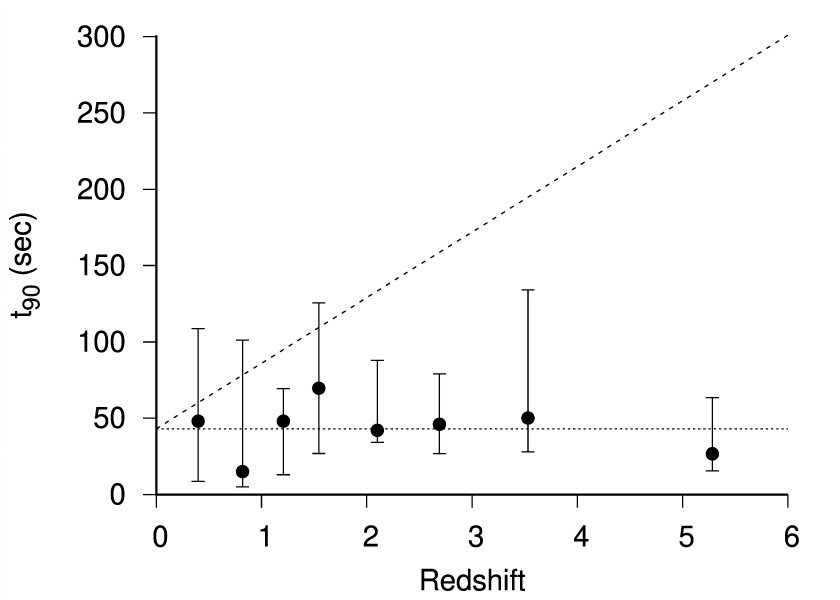}
\vskip -0.3 cm
\caption[]{Duration of gamma-ray bursts, as a function of redshift.
Each point (filled circle) is the median of the t$_{90}$ values of the 17 brightest GRBs, among 51 GRBs with similar redshifts (7 among 21, for the last point), error bars indicating the interquartiles.    
Horizontal dotted line: t$_{90}$=43 sec.
Dashed line: trend expected for the median value, according to metric theories of gravity.}
\label{Fig:dilation}
\end{figure}

\section{Discussion}

\subsection{Are tired-light models still relevant ?}

It has been claimed that theories where the Hubble-Lema\^itre law is explained by a loss of energy of the photons during their travel are excluded \cite{Sandage:01}, noteworthy because,
as predicted by Lema\^itre cosmologies \cite{Wilson:39},
the SN Ia light curves seem dilated by a (1 + z) factor.
 
However, it has also been claimed that there is no such time dilation in the light curves of quasars \cite{Hawkins:01,Hawkins:10} or in duration measures of
GRBs \cite{Petrosian:13,Levan:13,Butler:14,Crawford:18}, 
casting doubts on a key prediction of metric theories of gravity, namely, the generality of the phenomenon.
Indeed, as shown in Figure \ref{Fig:dilation}, the durations of GRBs (t$_{90}$) do not seem significantly different at low and high redshifts (up to $z \approx$ 5.3). Note that, in order to cope with the fact that the GRB signal can be noisy at high redshifts \cite{Tagliaferri:12,Petrosian:13}, only the brightest GRBs were retained for determining the median of the t$_{90}$ values at a given redshift. To do so, the full \textit{Swift} dataset considered above was sorted by increasing redshift values and split into eight groups of 51 GRBs with similar redshifts.\footnote{With 21 GRBs in the highest-redshift group.} Then, for each group, $t_{90}$ values of one third of the GRBs were considered, namely, those with the highest BAT 1-sec peak photon flux.

Interestingly, the average over the eight median $t_{90}$ values is 43 $\pm$ 15 sec, while at $z =$ 3.5, for instance, the median value predicted   
by metric theories of gravity is $t_{90} = 43 \times ( 1 + z ) \approx$ 200 sec, which is excluded at the 10$\sigma$ level.
    
In the context of the present study, the time dilation of SN Ia light curves is instead expected to be either the signature of some evolutionary process \cite{Wasserman:00,Ree:20,Perlmutter:20}, or due to cosmology-dependent assumptions made during the analyses of the SN Ia light curves \cite{Melia:13,Crawford:17}.

\subsection{How can the Universe be stationary ?}

The hypothesis that the Universe is stationary has been taken for granted for long\cite{Einstein:17}, being noteworthy put forward within the frame of steady-state cosmologies \cite{Bondi:48,Hoyle:48,Bondi,Narlikar:87}. In the later case, it was however in a different context. In particular, the space-time metric was \textit{not} considered as being static.

On the other hand, herein, stationarity is observed on galactic scales, namely, when number densities of galaxies are considered. This result suggests that there is a repulsive force able to cancel out gravitational attraction, when the distance between galaxies is of the order of magnitude of the average distance between neighboring ones. 
Note that, in order to counteract such an attraction, this force has to yield accelerations of the order of 10$^{-12}$ m$\cdot$s$^{-2}$, well below the lowest values measured on Earth \cite{Aspelmeyer:21} or in the solar system \cite{Anderson:98,Giudice:06}. 
Note also that stationarity can be observed only if this force has a distance dependence steeper than Newtonian gravitation, a criterion that is for instance not met by the force associated to the cosmological constant \cite{Eddington:30}. 

\subsection{How are photons lost ?}

With eqn \ref{eq:range} and eqn \ref{eq:cste},
eqn \ref{eq:opacity} can also be written as:
$$
\tau(z) = \frac{\Delta t}{\tau_\gamma}
$$
where $\tau_\gamma = \frac{1}{\varepsilon H_0}$ 
is the photon lifetime along the line of sight. So,
$\varepsilon$ = 3.1 means that after $\approx$ 3 Gyr of travel half of the photons of a quasar or of a SN Ia are missing. 

Absorption by massive amounts of dust could prove responsible for this loss.
However, quasar luminosity-distances analyzed herein 
were determined by comparing their X-ray and UV fluxes, 
as well as their fluxes in the optical,
so as to obtain precise estimates for a possible extinction.\cite{Lusso:19,Lusso:15}. 
SN Ia distance moduli were also determined in the optical,
so, such dust would have to be "grey" \cite{Hannestad:99,Cepa:07} over a three-decade frequency range, at least. 
On the other hand, numerous studies have shown
that dust in significant amounts can not be made of grains smaller than $\approx$ 1 $\mu m$,
because otherwise it would have been detected by its reddening,\cite{Wright:81,Aguirre:99,Corrales:15}
and also because grains of this size would scatter soft X-rays, producing a diffuse halo around X-ray point sources \cite{Paerels:12}.
Interestingly, it has been argued that 
during the ejection process from galaxies,
or within the intergalactic medium itself,
a preferential destruction of small grains could occur \cite{Aguirre:99}.
However, there are also severe observational constraints on the distribution of 
large amounts of dust around galaxies.
For instance,
little extinction has been found within the central parts of galaxy clusters,
meaning that no such dust colocalize 
with the gas in the intra-cluster medium \cite{Yamada:09,Corredoira:17}.

On the other side of the mass spectrum, 
the hypothesis that there may have 
large amounts of objects in the vicinity of galaxies with masses over $\approx$ 10$^{-7}$ M$_\odot$ seems to have been excluded by microlensing surveys, for objects with masses up to $\approx$ 30 M$_\odot$ \cite{Macho:98,Macho:01},
and by studies of their possible effects on the stability or on the dynamics of galaxies, for heavier ones \cite{Ostriker:85,Brandt:16}.
However, in the former case, it has been argued 
that the interpretation of the results strongly depends
upon the model parametrization chosen for the halo of the Milky Way.
\cite{Hawkins:15}.

On the other hand, it has been suggested that photons could have a finite lifetime \cite{Marchegiani:14,Heeck:13}, \textit{e.g.}, by decaying into
lighter particles such as massive neutrinos \cite{Hari:76,Pastor:06}, thus
reducing their flux along the line-of-sight.
This would noteworthy mean that photons have a small, yet nonzero rest mass m$_\gamma$ of the order of:
$$
m_\gamma \approx \frac{h}{\tau_\gamma c_0^2} \approx 10^{-68} \mathrm{kgs} 
$$ 
that is, well below
current upper limits \cite{Spallicci:16,Meszros:16}.
\vspace{-5pt}

\subsection{Can H$_0$ be measured on Earth ?}

Eqn \ref{eq:newhubblet} suggests that, like in the case of 
most tired-light models,
the Hubble constant could be measured in laboratory experiments, as a frequency drift proportional to the photon time-of-flight. Note however that such a measurement would be a challenging one, the expected drift being of the order of $10^{-18}$ s$^{-1}$. 

But note also that this frequency drift may not occur on distance scales that small.  
As a matter of fact, in the course of the present study, the photon lifetime along the line-of-sight has been found to be nearly one third of the Hubble time. 
Such a numerical coincidence may prove significant. It could for instance mean that there is a causal relationship between the way photons interact with galaxies and their cosmological frequency drift, as claimed long ago by Fritz Zwicky \cite{Zwicky:29}.

\section{Conclusion}

The present study shows that, by combining a tired-light model where $H(z) = H_0 ( 1 + z )$, in fair agreement with observational data (Fig. \ref{Fig:hofz}), with the hypothesis that the Universe is far from being as transparent as assumed nowadays \cite{Holanda:14,Dantas:17,Holanda:18}, it is possible to obtain a two-parameter luminosity distance (eqn \ref{eq:Dlgeneral}) able to match observations up to $z \approx 5$ (Fig. \ref{Fig:quasars} and \ref{Fig:snIa}). 

Interestingly, for $z \gg 0$, the corresponding cosmic distance duality relation (eqn \ref{eq:duality} and \ref{eq:eta_of_z}) differs from the prediction of metric theories of gravity. Moreover, at least for $z < 1$, it seems in fair agreement with observations (Fig. \ref{Fig:duality}).

In this context, as far as galaxy number densities are concerned, the Universe looks stationary, up to $z \approx$ 2 at least (Fig. \ref{Fig:count}). Also, up to $z \approx$ 4, the average linear size of galaxies does not seem to vary significantly, in sharp contrast with the dramatic evolution that has to be postulated in order to accommodate $\Lambda$CDM \cite{Vandokkum:17,Nakajima:19,Lopez:10}.

In order to turn such results into a consistent cosmology, 
a number of questions remain to be addressed, such as:
what is the physical origin of the current stability of the Universe on galactic scales ? What is the physical origin of the loss of energy of photons during their travel ?
A consistent alternative for the nucleosynthesis of light elements is also needed.   

At a more general level, the apparent lack of dilation of the durations of GRBs as a function of redshift\cite{Petrosian:13,Levan:13,Butler:14,Crawford:18} (Fig. \ref{Fig:dilation}) seems to rule out an explanation of the cosmological redshift based upon metric theories of gravity.

\section*{Acknowledgments}

I thank the referee for his careful and constructive reading of the manuscript.
I also thank Guido Risaliti, for providing the dataset of quasar luminosity-distances,  
Gabriel Chardin, 
Martin Lopez-Corredoira, Louis Marmet, Georges Paturel and Marcel Urban, for insightful comments.


\end{document}